\documentstyle[multicol,epsf,aps]{revtex}
\begin{document}
\title{Stochastic Aggregation: Scaling Properties}
\author{E.~Ben-Naim$\dag$ and P.~L.~Krapivsky$\ddag$}
\address{$\ddag$Theoretical Division and Center for Nonlinear Studies, 
Los Alamos National Laboratory, Los Alamos, NM 87545}
\address{$\dag$Center for Polymer Studies and Department of Physics,
Boston University, Boston, MA 02215}
\maketitle 
\begin{abstract} 
  We study scaling properties of stochastic aggregation processes in
  one dimension. Numerical simulations for both diffusive and
  ballistic transport show that the mass distribution is characterized
  by two independent nontrivial exponents corresponding to the
  survival probability of particles and monomers.  The overall
  behavior agrees qualitatively with the mean-field theory.  This
  theory also provides a useful approximation for the decay exponents,
  as well as the limiting mass distribution.
\end{abstract}
\begin{multicols}{2}
\section{Introduction}

In the previous study, we introduced a stochastic aggregation process
involving both active and passive clusters\cite{prec}.  We generalized
Smoluchowski's rate equations and obtained exact results for several 
kernels.  In this study, we apply stochastic aggregation to
reaction-diffusion, coarsening, and ballistic agglomeration problems.
Our goal is to examine the range of validity the mean-field results,
and to determine whether the overall scaling behavior extends to low
dimensional systems.

The rate equations approach is mean-field in nature, i.e., it is valid
only when spatial correlations are absent.  Formally, it is applicable
only in infinite spatial dimension, or in the presence of an effective
mixing mechanism. This mean-field theory should also be {\em
  asymptotically} exact when the spatial dimension is sufficiently
high. In low spatial dimensions, however, significant spatial
correlations eventually develop, and the rate equation approach does
not apply in the long time limit. We therefore focus on
one-dimensional systems where spatial correlations are most
pronounced.

We performed numerical simulations of stochastic aggregation processes
with both diffusive and ballistic particle transport.  The simulations
show that the scaling behavior suggested by the mean-field theory is
indeed generic, as it extends to one-dimensional systems.  We find
that two nontrivial model-dependent exponents characterize the
survival probabilities of the particles and monomers, respectively.
Smoluchowski's theory provides reasonable estimates for these exponents.

Additionally, we studied the limiting mass distribution of passive
clusters.  Surprisingly, over a substantial mass range, this
distribution depends only weakly on the underlying transport
mechanism.  Furthermore, mean-field theory provides an excellent
approximation for the limiting mass distribution.

The rest of this paper is organized as follows. The general scaling
behavior is outlined in Sec.~II. Predictions of the mean-field theory
are summarized in Sec.~III. Numerical simulations of stochastic
aggregation processes with diffusive and ballistic transport
mechanisms are described in Secs.~IV and V, respectively.  A
discussion of the results is presented in Sec.~VI.

\section{Scaling Properties}

Stochastic aggregation involves two types of clusters: active and
passive\cite{prec}. Initially, the system consists of active monomers
only. When two active clusters merge, the newly-born aggregate remains
active with probability $p$, or becomes passive (i.e., it never
aggregates again) with probability $q=1-p$. Eventually, all active
clusters are depleted and the system consists of passive clusters
only.  This process can be viewed as an aggregation-annihilation
process since it interpolates between aggregation ($p=1$) and
annihilation ($p=0$)\cite{aa}.

Quantities of interest include $A_k(t)$ and $P_k(t)$, the
distributions of active and passive clusters at time $t$, as well as
the final distribution of passive clusters, $P_k(\infty)$. As shown in
\cite{prec}, two conservation laws underly this system. The first is
mass conservation, $\sum k[A_k(t)+P_k(t)]={\rm const}$.  The second
conservation law reflects the fact that changes in the overall
densities are coupled, $qA(t)+(1+q)P(t)={\rm const}$, where $A(t)=\sum
A_k(t)$ and $P(t)=\sum P_k(t)$ are the number densities of
active and passive clusters, respectively.

Therefore, it is sufficient to study the time evolution of the number
density and the mass density of active clusters, $A(t)$ and $M(t)=\sum
kA_k(t)$, respectively. The latter quantity is the survival
probability of an active particle, i.e., the probability that it still
belongs to an active cluster at time $t$.  Smoluchowski's theory
suggests that both quantities decay algebraically in the long time
limit
\begin{equation}
\label{rs}
A(t)\sim t^{-\nu}, \qquad M(t)\sim t^{-\nu\psi}.
\end{equation}
As will be shown below, this as well as other scaling properties
suggested by this theory hold qualitatively even for low dimensional
stochastic aggregation processes.  While the decay exponent $\nu$ is
typically robust in that it depends only on the major characteristics
of the process such as the spatial dimension or the transport
mechanism, the exponent $\psi\equiv \psi(p)$ is non-universal as it
depends on the details of the model, i.e., on the probability $p$. In
turn, this implies a non-universal growth law for the average mass of
an active cluster $\langle k\rangle=M/A\sim t^{\nu(1-\psi)}$.

For the system to follow a scaling behavior, the average mass must be
the only relevant scale in the long time limit, and conversely, any
scale characterizing the initial mass distribution must be ``erased''
eventually. In other words, the mass distribution is characterized by a
single rescaled variable 
\begin{equation}
\label{rk}
A_k(t)\sim t^{\nu(\psi-2)}F\left(kt^{\nu(\psi-1)}\right), 
\end{equation}
with the time dependent prefactor fixed by the decay laws
(\ref{rs}).

This scaling behavior is similar to that found for deterministic
aggregation-annihilation processes\cite{k,bk95,yang} and for
aggregation-annihilation of domains in coarsening
processes\cite{kb,bk,lm}.  These studies suggest that another
independent exponent describes the decay of small clusters.
Specifically, the monomer density decays according to
\begin{equation}  
\label{r1}
A_1(t)\sim t^{-\nu\delta},
\end{equation}
with a model-dependent exponent $\delta\equiv\delta(p)$.  The monomer
density decay reflects the small argument behavior of the scaling
function $F(\xi)\sim \xi^{\sigma}$ with $\delta-1=(1-\psi)(1+\sigma)$.
One of our main results is that the mass distribution of active
clusters is described by a set of nontrivial exponents $(\psi,\delta)$.
These exponents can be viewed as persistence
exponents\cite{dbg,kbr} as they characterize the survival probability
of an active particle, and an active monomer \cite{mc}.

Several properties of the scaling exponents are general. For instance,
the inequalities $\psi\leq 1\leq\delta$ hold since $A_1\leq\sum
A_k\leq\sum kA_k$. The two exponents are equal $\psi=\delta=1$ in the
annihilation case ($p=0$), since $A_k(t)=A(t)\delta_{k,1}$.  In the
aggregation limit ($p=1$) the mass density of active clusters is
conserved and therefore $\psi=0$.

We now turn to the mass distribution of passive clusters. The
Smoluchowski theory suggests that the same scaling form underlies both
mass distributions 
\begin{equation}
\label{pk}
P_k(t)\sim t^{\nu(\psi-2)}G\left(kt^{\nu(\psi-1)}\right). 
\end{equation}
In contrast with the active cluster distribution, the passive cluster
distribution approaches a nontrivial final distribution $P_k(\infty)$.
Such a {\em time independent} final distribution is consistent with
the above scaling form only when the scaling function diverges,
$F(\xi)\sim \xi^{-\gamma}$ in the limit $\xi\to 0$, with
$\gamma=(2-\psi)/(1-\psi)$. As a result, the final mass distribution
of passive clusters decays algebraically in the large mass limit
\begin{equation}
\label{pkg}
P_k(\infty)\sim k^{-\gamma} \quad {\rm with}\quad 
\gamma={2-\psi\over 1-\psi}.
\end{equation} 
At a given time $t$, this decay is realized for clusters whose mass
$k$ does not exceed the characteristic mass $k^*\sim t^{\nu(1-\psi)}$. 
Note also that $0<\psi<1$ implies $2<\gamma<\infty$.  Generally, the
mass conservation restricts the large mass decay exponent to
$\gamma>2$. Since the $\psi$ exponent varies between $0$ and $1$, we
see that the entire range of acceptable exponents is realized by
tuning the probability $p$.

\section{Mean-Field Theory}

It is well established that spatial correlations can be safely neglected
only in spatial dimensions larger than some upper critical dimension,
$d>d_c$\cite{sid}.  For example, for irreversible aggregation with
mass-independent diffusion and reaction rates, one has $d_c=2$; for a
general aggregation process, however, the upper critical dimension may
be arbitrarily large\cite{van}. Below the upper critical dimension,
substantial spatial correlations develop, and the most important
features including the scaling exponents and the scaling functions are
changed. Generally, the lower the spatial dimension, the larger the
difference with the mean-field predictions.

Although the Smoluchowski rate equations approach does not apply in
low spatial dimensions, it can still serve as a useful approximation
after an appropriate modification.  This can be accomplished by
replacing the overall reaction rate with an effective 
density-dependent reaction rate $r\equiv r(A)$ 
\begin{eqnarray}
\label{mft}
{d A_k\over dt}&=&r \left({p\over 2}\sum_{i+j=k}A_i A_j
-A_kA\right),\nonumber\\
{d P_k\over dt}&=&r \left({q\over 2}\sum_{i+j=k}A_iA_j\right).
\end{eqnarray}
We are primarily interested in situations where aggregation is
independent of the mass, and therefore we use a mass independent rate
kernel.  The reaction rate $r(A)$ is model dependent.  In
reaction-diffusion processes, the reaction rate decays algebraically
with the density (see, e.g., Refs.\cite{bid,be}).  Assuming $r(A)\sim
A^{\alpha}$ yields ${dA\over dt}\sim -A^{\alpha+2}$, and consequently,
the density decay exponent is found
\begin{eqnarray}
\label{nu}
\nu={1\over 1+\alpha}.
\end{eqnarray}
In general, a reduction in the reaction rate, i.e., $\alpha>0$, leads
to a slowing down in the density decay rate, $\nu<1$.  Apart from the
change in $\nu$, all other aspects of this approximation are identical
to the Smoluchowski theory with a constant rate kernel. Indeed, the
above rate equations reduce to the Smoluchowski's rate equations with
a redefined time variable, $t\to \tau=\int_0^t dt'\, r(t')$. In
particular, the scaling exponents $\psi$ and $\delta$ are
independent of $\alpha$:
\begin{equation}
\label{pd}
\psi=2\,{1-p\over 2-p},\qquad \delta={2\over 2-p}.
\end{equation}
One can verify the expected limiting behaviors $\psi(1)=0$, and
$\psi(0)=\delta(0)=1$.  Furthermore, the scaling functions are as in
the constant kernel solution \cite{prec}, and for example, $F(\xi)$ is
purely exponential.  The corresponding small argument exponents
$\gamma=2/p$ and $\sigma=0$ follow from $\psi$ and $\delta$ using the
aforementioned scaling relations.  The final mass distribution of
passive clusters is  independent of the reaction rate $r$ \cite{prec}
\begin{equation} 
\label{pkinf}
P_k(\infty)={q\over p}\,{\Gamma(1+2/p)\,\Gamma(k)\over \Gamma(k+2/p)}.
\end{equation}

Below, we compare these mean-field predictions with simulation results
for one-dimensional stochastic aggregation where spatial correlations
are most pronounced. We also examine the role of the aggregates'
transport mechanism by considering both diffusive and ballistic 
transport.

\section{Diffusive Transport}

In diffusive stochastic aggregation, identical particles are placed onto
a $d$-dimensional lattice. All particles perform independent random
walks, i.e., they hop to a randomly chosen nearest-neighbor site with a
constant rate. If this site is occupied, the two particles coalesce
irreversibly, and with probability $p$ the resulting aggregate remains
active, while with probability $q=1-p$ it becomes passive.  Effectively,
passive particles are removed from the system.

In the case of single-species reaction diffusion processes, the
effective reaction rate can be obtained from dimensional analysis.
Eq.~(\ref{mft}) implies $[r]=[L]^d[T]^{-1}$, and since the reaction rate
can only be a function of the diffusion coefficient $[D]=[L]^2[T]^{-1}$
and the density $[A]=[L]^{-d}$, one finds $r\propto DA^{(2-d)/d}$.
Hence, $\alpha=(2-d)/d$ and Eq.~(\ref{nu}) yields the correct decay
exponents $\nu=d/2$\cite{sid} below the upper critical dimension
$d_c=2$.

To examine the above scaling picture we performed numerical simulations
of diffusive stochastic aggregation processes in one dimension.  Unless
noted otherwise, the data was obtained from an average over 10
independent realizations in a system of size $L=10^7$ with periodic
boundary conditions. Initially, all sites were occupied.  First, we
verified that the number density, the mass density, and the monomer
density indeed decay algebraically in the long time limit, in accord
with Eqs.~(\ref{rs}) and (\ref{r1}).  The case $p=1/2$ is shown in
Fig.~1, and the corresponding decay exponents were found:
$\nu=0.500(1)$, $\psi=0.6193(3)$, and $\delta=1.460(2)$. Mean-field
theory correctly predicts $\nu=1/2$.  Furthermore, the predictions
$\psi=2/3$ and $\delta=4/3$ provide a reasonable approximation. One can
compare with the case of disordered (Sinai) diffusion where a real-space
decimation procedure\cite{fisher} was used to determine {\em exact}
values of these exponents\cite{lm}.  Remarkably, the disorder case
exponent $\psi=0.61937$ is in excellent agreement with the simulation
value. There is a small discrepancy with the second exponent
$\delta=1.47041$. Additionally, we verified that the densities of active
and passive clusters follow the scaling forms of Eqs.~(\ref{rk}) and
(\ref{pk}), respectively (see Fig.~2). In agreement with the mean-field
theory, the scaling functions decay exponentially for large masses.

\begin{figure}
\centerline{\epsfxsize=7cm \epsfbox{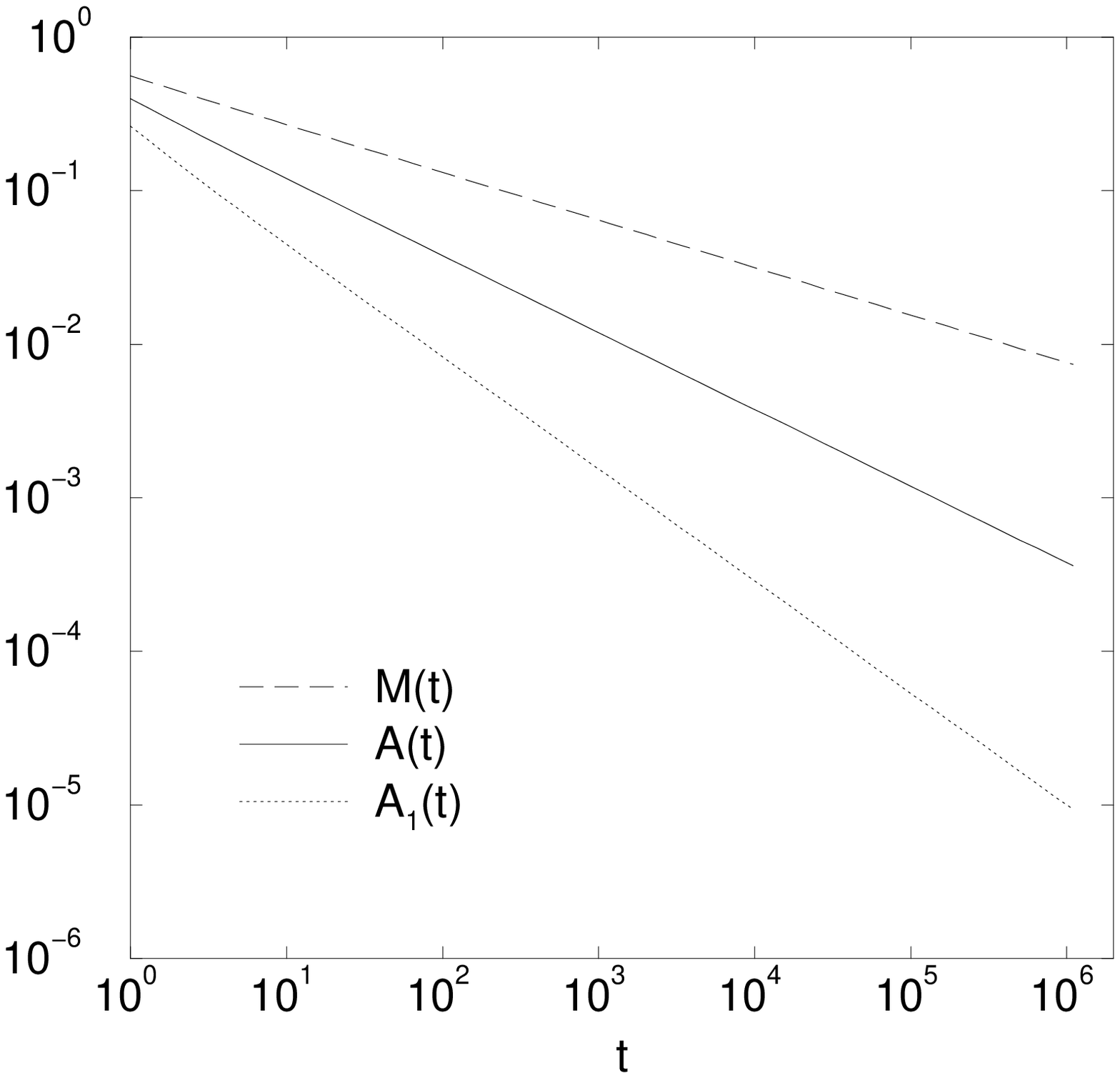}}
\noindent{\small {\bf Fig.1} The  number density, the  
  mass density, and the monomer density versus time for
   $p=1/2$.}
\end{figure}

\begin{figure}
\centerline{\epsfxsize=7.5cm \epsfbox{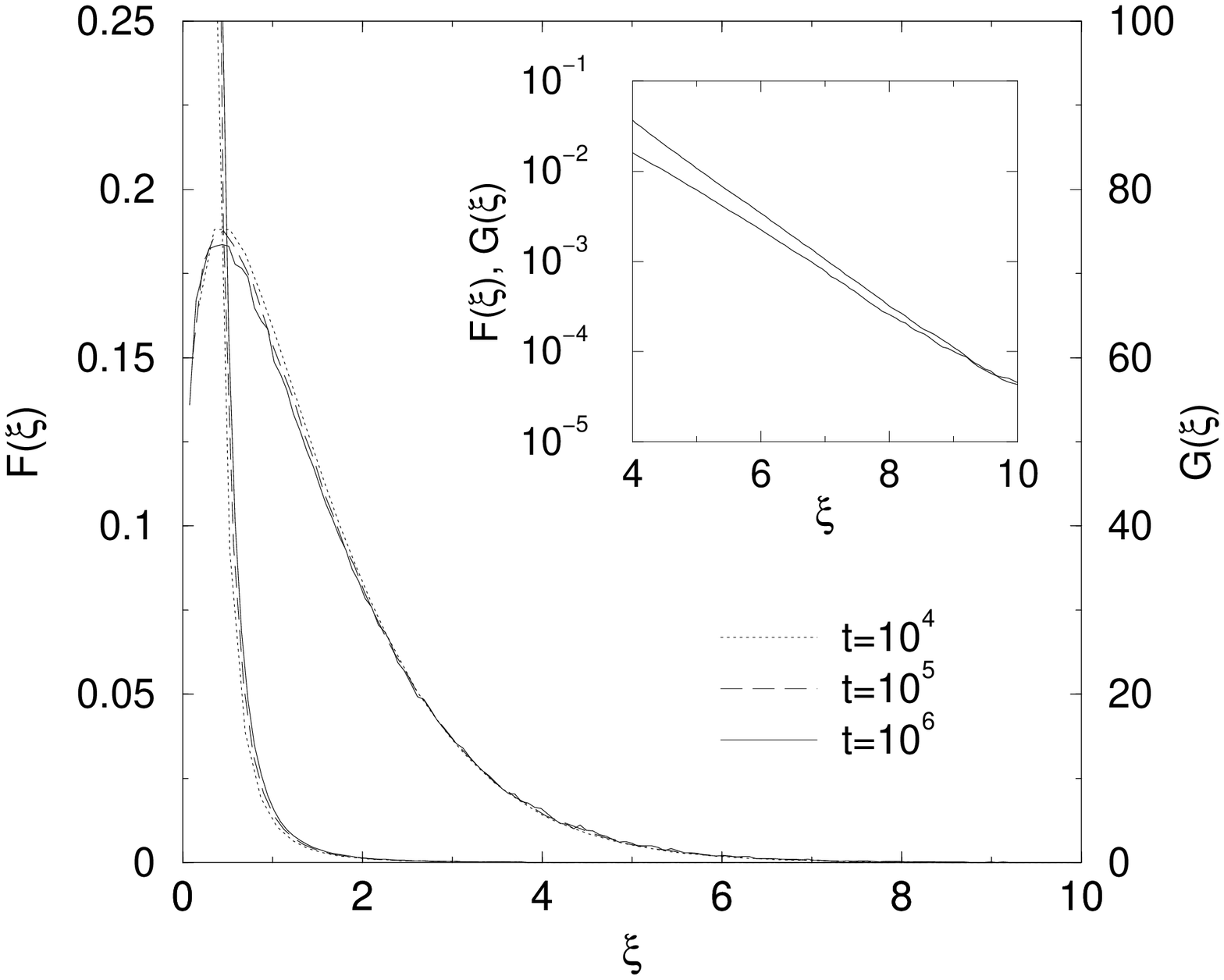}}
\noindent{\small {\bf Fig.2} Scaling of the active and passive mass 
  distributions. Shown are the scaling functions $F(\xi)\equiv
  t^{\nu(2-\psi)}A_k(t)$, and $G(\xi)\equiv t^{\nu(2-\psi)}P_k(t)$,
  versus the scaling variable $\xi=kt^{\nu(1-\psi)}$ at three 
  different times $t=10^4$, $10^5$, and $10^6$. Different scales
  correspond to $F(\xi)$ and $G(\xi)$ in the main figure since the
  latter diverges at the origin.  The data represent an average over
  $10^3$ independent realizations in a system of size $L=10^6$ for the
  case $p=1/2$. The exponent value $\psi=0.619$ was used. The tail of
  the distribution is shown in the inset.}
\end{figure}
We also studied how the exponents vary with the probability $p$, as
shown in Figs.~3 and 4.  The exact exponents found for the disordered
case by Le Doussal and Monthus\cite{lm} provide an excellent
approximation (within 0.1\%) for $\psi$.  In the case of $\delta$, a
different behavior emerges in the aggregation limit, $p\to 1$, where the
exact value is $\delta=3$ \cite{sp}, and the disagreement with both
mean-field theory and the disordered case are most pronounced. 

\begin{figure}
\centerline{\epsfxsize=7cm \epsfbox{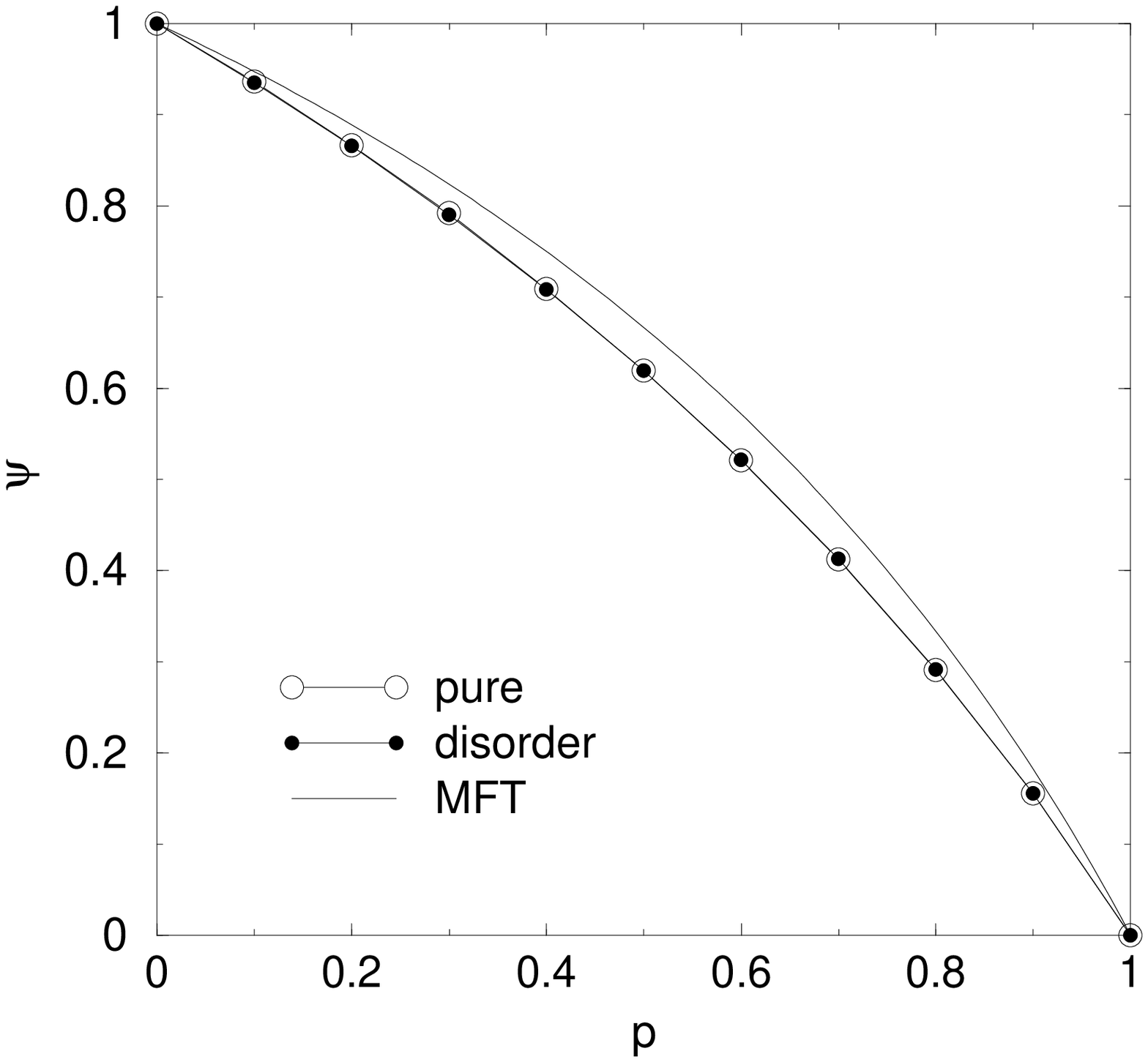}}
\noindent{\small {\bf Fig.3} The exponent $\psi$ 
  versus $p$. Monte Carlo simulation results for the pure diffusion
  case are compared with the mean-field theory (\ref{pd}) and the
  exact value for the disordered case. The latter is obtained from 
  ${}_2F_0(-{2\over 2-p},2\psi,2)=0$\cite{lm},  where ${}_2F_0(a,b,z)$
  is the confluent hypergeometric function.}
\end{figure}

\begin{figure}
\centerline{\epsfxsize=7cm \epsfbox{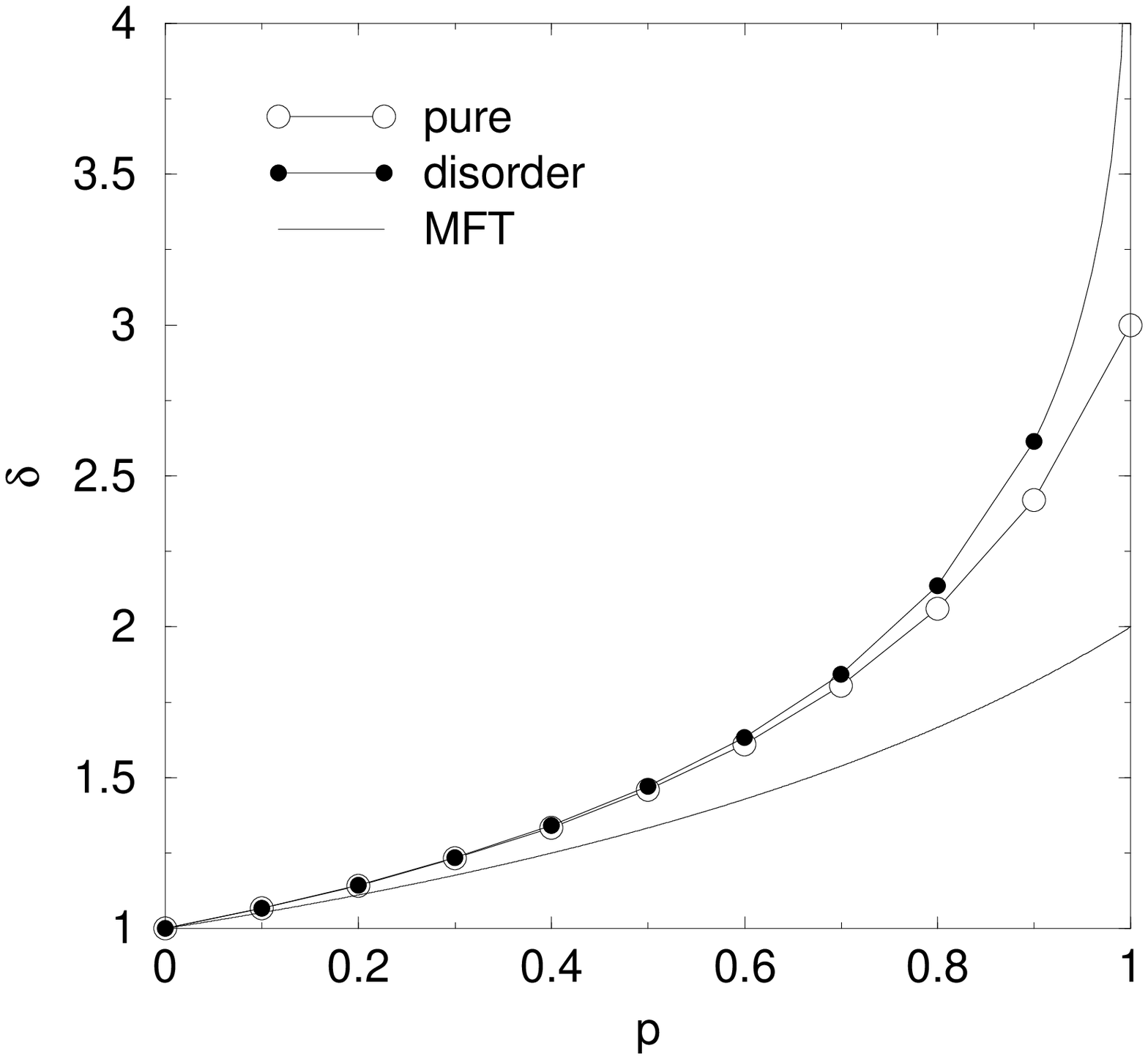}}
\noindent{\small {\bf Fig.4}
   The exponent $\delta$ versus $p$. Monte Carlo simulation results for
   the pure diffusion case are compared with the mean-field theory
   (\ref{pd}) and the exact value for the disordered case
   obtained from ${}_2F_0(-2{1-p\over 2-p},2\delta,2)=0$\cite{lm}.}
\end{figure}

The above scaling arguments suggest that the limiting mass
distribution of passive clusters decays algebraically with the
exponent $\gamma=(2-\psi)/(1-\psi)$.  For $p=1/2$, one therefore
expects $\gamma\cong 3.627$ (compare with $\gamma=3.62722$ and
$\gamma=4$, predicted by the disordered case and the mean-field
theory). This corresponds to a very strong suppression of large
masses, and therefore, it is much more difficult to confirm this
behavior numerically.  Nevertheless, our simulations (Fig.~5) are
consistent with a power law decay with an exponent $\gamma\cong 3.6$.

In one dimension, the diffusion-controlled stochastic aggregation is
equivalent to the Potts model with zero-temperature Glauber
dynamics\cite{gl}.  For the $Q$-state Potts model with spatially
uncorrelated initial conditions, aggregation of domain walls occurs
with probability $p={Q-2\over Q-1}$, and annihilation occurs with
probability $q={1\over Q-1}$.  Therefore, the above can be
reformulated in terms of domain walls rather than aggregates. In the
coarsening context, the domain wall mass (or number) distribution is
dual to domain number distribution \cite{kb,bk,lm}.

\begin{figure}
\centerline{\epsfxsize=7cm \epsfbox{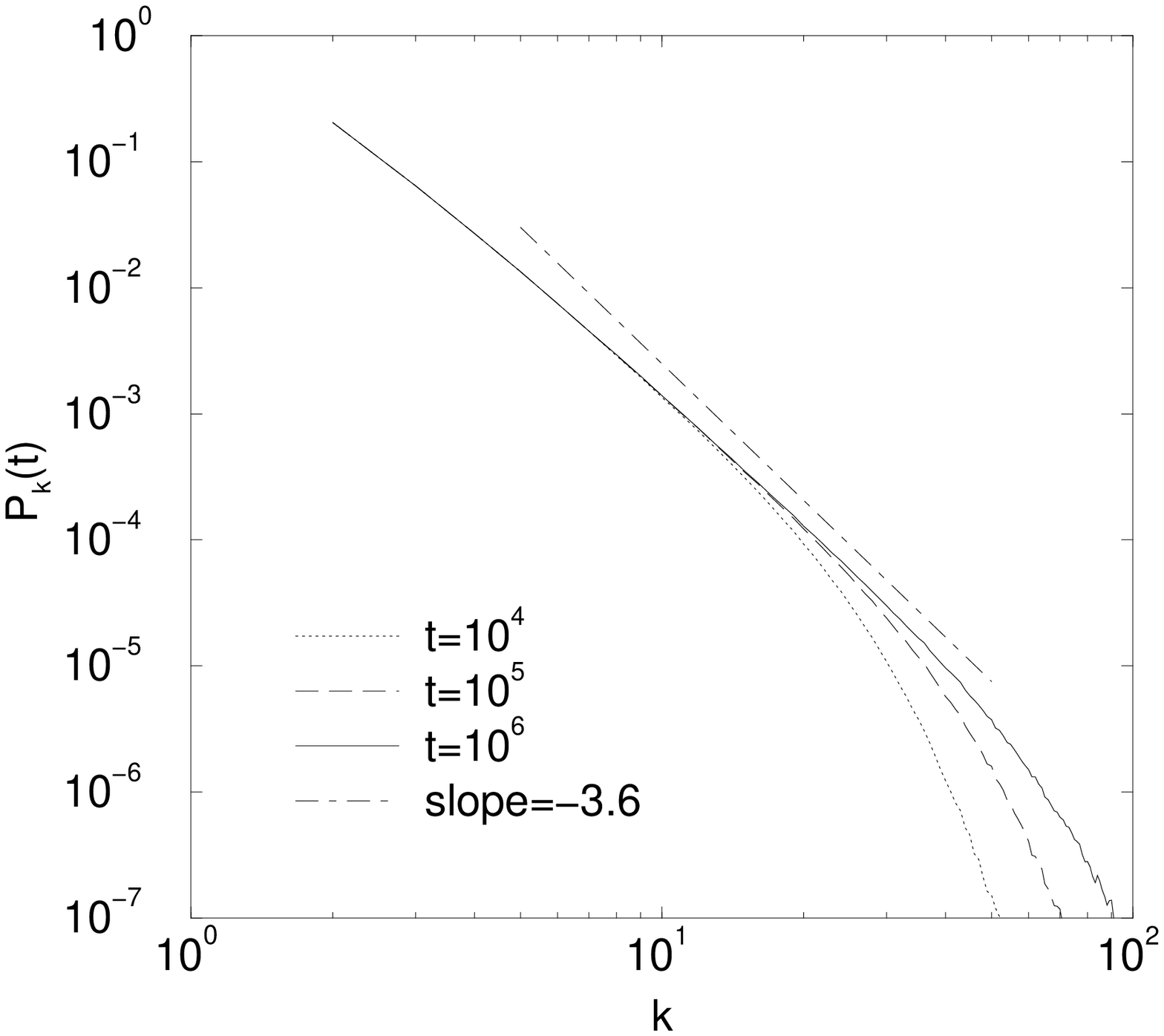}}
\noindent{\small {\bf Fig.5} The distribution $P_k(t)$ versus $k$ for three
   different times $t=10^4$, $10^5$ and $10^6$. The typical mass at
   these three times is proportional to $k^*\equiv
   t^{\nu(1-\psi)}\propto 6$, $10$, and $16$, respectively.  Hence, the
     distribution is fully developed only over a short range of masses.
     The data represents an average over $10^3$ realizations in a
     system of size $L=10^6$ with $p=1/2$.}
\end{figure}

\section{Ballistic Transport}

The situation when particles move ballistically involves several
complications.  First, while the annihilation limit is uniquely
defined \cite{ef,brl,krl,drfp,rdp}, the aggregation limit has various
realizations. In traffic flows, the velocity of a newly-born cluster
is the smaller of the two velocities \cite{bkr}, while in application
to astrophysics and granular gases the velocity follows from momentum
conservation \cite{sz,bcdr}.  Second, numerical results for the
annihilation case \cite{brl} and analytical results for the traffic
case \cite{bkr} show that the initial conditions are remembered
forever, in contrast with the diffusive case. Specifically, the small
velocity characteristics of the initial velocity distribution
influence the long time asymptotic behavior, including the scaling
exponents. 

We consider the momentum conserving case, also known as ``ballistic
aggregation'' or ``sticky gas'' \cite{cpy,pi,jl,mp,kb96,f,fmp}.  The
initial velocities are assigned according to the distribution
$P_0(v)$.  The mass (momentum) of a newly-born cluster is equal to
the sum of masses (momenta) of the two colliding clusters.  After an
agglomeration event, the newborn particle remains active with
probability $p$, or becomes passive with probability $q=1-p$.

To apply the Smoluchowski rate equations approach, we again use
dimensional analysis to calculate the decay exponent $\nu$. The
collision rate is $r\sim va^{d-1}$, where $v$ is the typical velocity
and $a$ is the typical radius of an aggregate.  A particle of radius
$a$ contains of the order $a^d$ monomers whose initial momenta are
uncorrelated. Momentum conservation therefore implies $v\sim
a^{-d/2}$. Using $a^d\sim M/A\sim A^{\psi-1}$ gives the collision
rate $r\sim a^{(d-2)/2}\sim A^{(d-2)(\psi-1)/2d}$.  From
Eq.~(\ref{nu}) one finds
\begin{equation}
\label{nub}
\nu={2d\over d+2+\psi(d-2)},
\end{equation}
with $\psi$ given by Eq.~(\ref{pd}). Apart from the exponent $\nu$,
features such as the exponential mass distribution and the exponents
$\psi$ and $\delta$ are given by the mean-field theory outlined above.
In two dimensions, the collision rate does not depend on $\psi$ and
hence, the asymptotic behavior $A\sim t^{-1}$ agrees with that found
for deterministic ballistic agglomeration\cite{cpy}. For $d\ne 2$,
stochastic and deterministic asymptotics differ: stochasticity
enhances decay of the number density $A$ for $d<2$ and weakens it for
$d>2$.  A more detailed mean-field theory can be carried. It yields a
factorizing joint mass-velocity distribution, with an exponential mass
distribution, and a Gaussian velocity distribution \cite{pi,kb96}.

In the aggregation case, $\psi=0$ and therefore the correct scaling
exponent $\nu=2d/(d+2)$ \cite{cpy} is recovered from Eq.~(\ref{nub}).
For the annihilation case, however, initial conditions are
``remembered'' forever and therefore the above dimensional arguments no
longer hold.  The predicted exponent in the annihilation case is always
mean-field $\nu=1$, while one-dimensional numerical simulations yield an
exponent continuously varying from $0$ to $1$ depending on the initial
velocity distribution $P_0(v)$, e.g., $\nu\approx 0.8$ for  uniform
initial distributions\cite{brl,rdp}.

We have simulated the stochastic aggregation process on a
one-dimensional ring with $10^6$ particles. The initial velocity
distribution was uniform in $[-1,1]$. We measured the scaling exponent
$\psi$ via the scaling relation $M\sim A^\psi$, rather than directly
versus time, since the exponent $\nu(p)$ is not known analytically.  We
have found that the mean-field prediction, $\psi=(2-2p)/(2-p)$, provides
a reasonable approximation for the exponent $\psi$, as shown in Fig.~6.
Furthermore, this approximation should improve in higher dimensions.

We compared the mean-field prediction for the mass distribution of
passive clusters, Eq.~(\ref{pkinf}), with the numerically obtained
distributions in both ballistic and diffusive cases. Interestingly,
the rate equations provide an excellent approximation for small and
moderate masses (see Fig.~7).  Given that the discrepancy in $\psi$ is
maximal for the case $p=1/2$, one may expect an even better
approximation for other values of $p$. Noting the strong decay of this
distribution, the contribution of very large masses is extremely
small; for example, $P_{100}(\infty)\approx 2.4 \times 10^{-7}$ for
$p=1/2$. Hence, the most pronounced part of the distribution is well
approximated by the rate equations theory.  Surprisingly, the
transport mechanism does not play an important role as far as the
final mass distribution of passive clusters is concerned.

\begin{figure}
\centerline{\epsfxsize=7cm \epsfbox{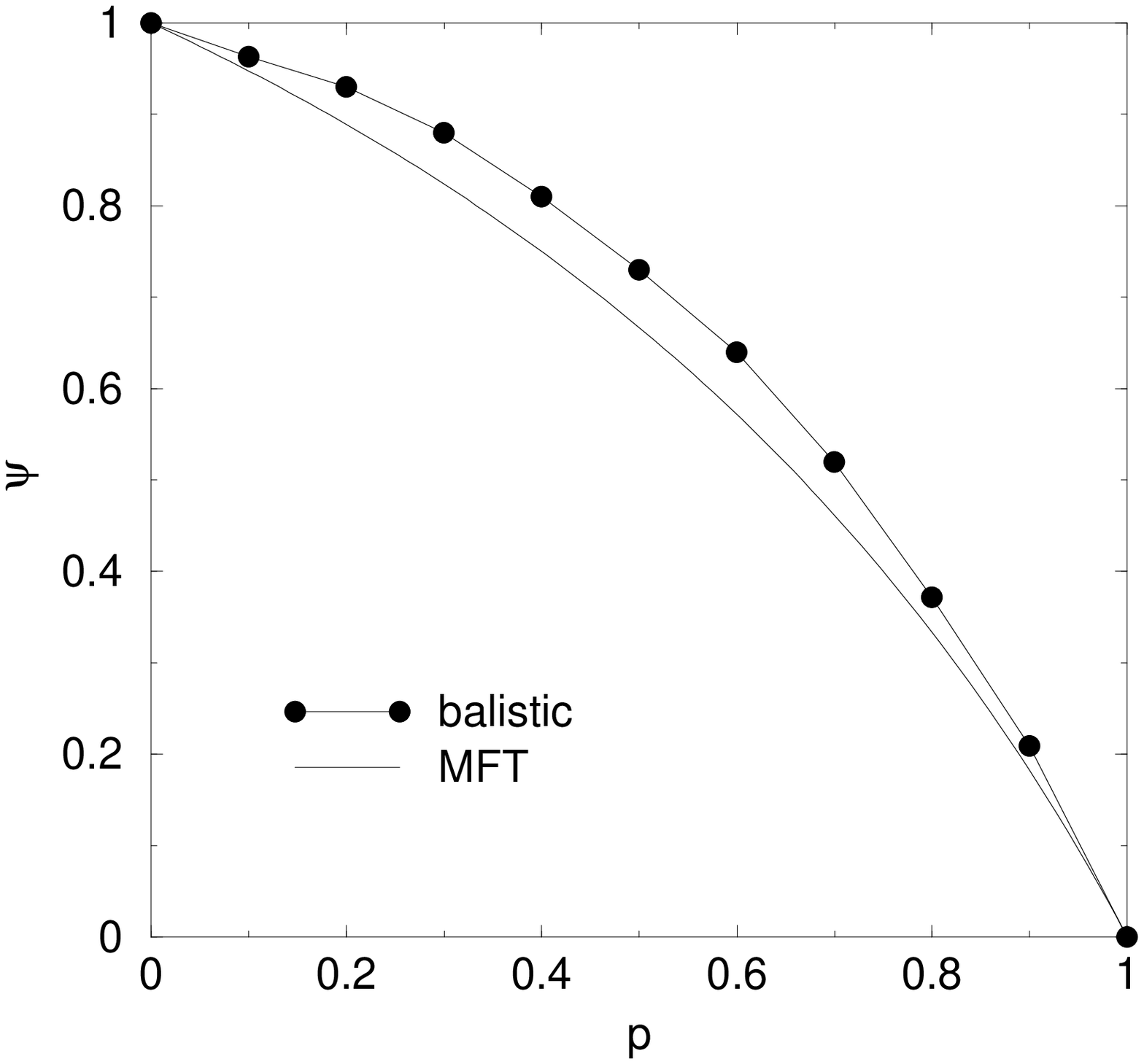}}
\noindent{\small {\bf Fig.6} The scaling exponent $\psi(p)$ versus $p$ for 
  ballistic aggregation compared with the mean-field value of Eq.~(\ref{pd}).}
\end{figure}

\begin{figure}
\centerline{\epsfxsize=7cm \epsfbox{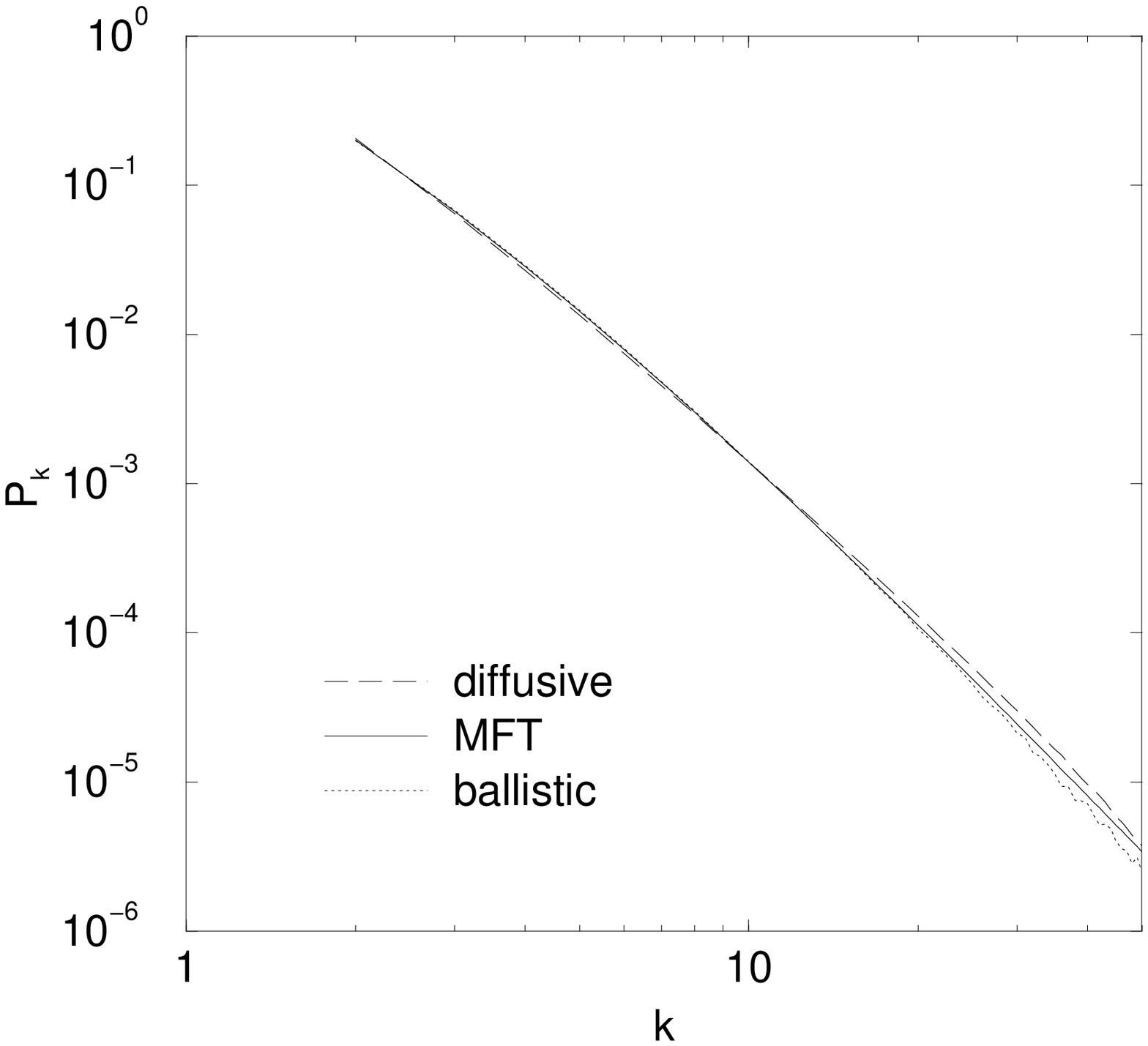}}
\noindent{\small {\bf Fig.7} The final distribution of passive 
  clusters for the $p=1/2$ stochastic aggregation with diffusive and
  ballistic transport. Also Shown is the mean-field distribution
  $P_k(\infty)=24/[k(k+1)(k+2)(k+3)$].}
\end{figure}

\section{Discussion}

We have investigated diffusion- and ballistic-controlled stochastic
aggregation in one dimension. We have seen that the rate equations
approach captures the overall scaling behavior and additionally it
provides reasonable estimates for the decay exponents.  In general,
the mass distribution is characterized by two nontrivial
model-dependent decay exponents.

In the diffusion-controlled case, the exponent $\psi$ underlying the
survival probability of a particle is in excellent agreement with the
exact results from the disordered case. In fact, one cannot dismiss the
possibility that the disordered and the pure values are identical, based
on numerics alone.  However, there is an evident discrepancy in the
exponent $\delta$ as the disordered case exponent diverges logarithmically
in the aggregation limit. Stochastic aggregation is equivalent to domain
coarsening in the zero-temperature Potts-Glauber model.  The above
exponents $(\psi,\delta)$ characterize the domain wall number
distribution in analogy with $(\psi_D,\delta_D)$ for the domain number
distribution \cite{kb}. In the latter case as well, exact values calculated
for the disordered case provide an excellent approximation for the
domain exponents. In general, the particle survival probability exponent
$\psi$ is robust, while the monomer survival probability exponent
$\delta$ is very sensitive to the details of the process.

In the ballistic-controlled case, we have shown that even in the
absence of a consistent mean-field theory, some characteristics such as
the exponent $\psi$ are well approximated by the rate equations.
Understanding of reaction processes with an underlying ballistic
transport remains largely incomplete. The asymptotic behavior is
highly sensitive to the initial conditions, and the critical dimension
is apparently infinite. In fact, exact analytical results are
available mostly in the aggregation limit \cite{bkr,f,fmp}.

The most intriguing property of the stochastic aggregation is the
profound lack of universality.  Indeed, the weak dependence on the
transport mechanism is in contrast with the strong dependence on the
parameter $p$.  For example, our numerical results show that the final
distribution of passive clusters is very close in diffusion- and
ballistic-controlled situations.  Another very impressive
manifestation of this is the excellent agreement between the values of
the exponent $\psi(p)$ in the disordered and pure cases.

\medskip\noindent 
This research was supported by the DOE
(W-7405-ENG-36), NSF (DMR9632059), and ARO (DAAH04-96-1-0114).

\end{multicols}
\end{document}